\documentstyle[aps,preprint,epsf,tighten]{revtex}

\begin{document}


\begin{titlepage}

\title{Pendulum Mode Thermal Noise in Advanced Interferometers: A comparison
of Fused Silica Fibers and Ribbons in the Presence of Surface Loss}

\author{Andri~M.~Gretarsson,\footnote{email: andri@phy.syr.edu}
Gregory~M.~Harry, 
Steven~D.~Penn,
Peter~R.~Saulson, 
William~J.~Startin.}
\address{Department of Physics, Syracuse University, Syracuse, NY 13244-
1130, USA}
\author{Sheila~Rowan}
\address{Ginzton Laboratory, Stanford University, Stanford, CA 94305-4085, USA}
\author{Gianpietro~Cagnoli, Jim~Hough}
\address{Department of Physics and Astronomy, University of Glasgow,
Glasgow, G12 8QQ, UK}

\maketitle


\begin{abstract}
The use of fused-silica ribbons as suspensions in gravitational
wave interferometers can result in significant improvements in
pendulum mode thermal noise.  Surface loss sets a lower bound to
the level of noise achievable, at what level depends on the
dissipation depth and other physical parameters. For LIGO II, the
high breaking strength of pristine fused silica filaments, the
correct choice of ribbon aspect ratio (to minimize thermoelastic
damping), and low dissipation depth combined with the other
achievable parameters can reduce the pendulum mode thermal noise
in a ribbon suspension well below the radiation pressure noise.
Despite producing higher levels of pendulum mode thermal noise,
cylindrical fiber suspensions provide an acceptable alternative
for LIGO II, should unforeseen problems with ribbon suspensions
arise.
\newline
PACS - 04.80.Nn, 95.55.Ym, 05.40.Ca
\end{abstract}

\thispagestyle{empty}

\end{titlepage}


\section{Introduction}
\label{sec:intro} One of the most important limitations to the
sensitivity of long baseline gravitational wave detectors such as
LIGO~\cite{Abramovici}, VIRGO~\cite{VIRGO}, GEO 600~\cite{GEO},and
TAMA~\cite{tama} is thermal noise associated with the test masses
and their suspensions. Designs for advanced detectors propose
either fused silica or sapphire test-masses. For fused silica
test-masses internal mode thermal noise is expected to be an
important source of noise from approximately 20 Hz to a few
hundred Hertz, whereas pendulum mode thermal noise is more
important below this range.

Pendulum mode thermal noise is due primarily to dissipation in the
suspending filaments.  It is imperative, therefore, to minimize
the intrinsic losses in the filament.  Many current detector
designs use metal wires to suspend the test-mass, but metals,
with $\phi \geq 10^{-5}$, result in unacceptably high levels of
pendulum mode thermal noise.  Fused silica has lower loss
($\phi\approx 3\times 10^{-8}$) and monolithic fused silica
suspensions have been shown to have much higher $Q$ than metal
wire suspensions~\cite{Braginsky,Lunin,Andri}. Such a monolithic
suspension system is being developed and adopted for use in the
GEO 600 detector~\cite{Rowan}, while variations of this design
are being considered for LIGO II.  In particular, fibers with
circular cross sections may be replaced with fused silica
ribbons~\cite{Rowan,Logan,white_paper}, which allow the suspension
filaments to be very thin and compliant in the direction of
motion~\cite{Weiss,Martin,Ju}.

Experiments indicate that in thin fused silica filaments much of
the dissipation takes place in a layer near the
surface~\cite{Andri}. The level at which surface loss affects the
total pendulum mode dissipation depends on the filament thickness
and geometry, and influences the choice of suspension design
parameters. To investigate this, we calculate the pendulum mode
thermal noise, including surface dependent loss, as a function of
the design parameters for fibers and for ribbons.


\section{Dissipation Dilution}
\label{diss_dil_sec}
In the absence of external sources of dissipation, the dissipation in the
fundamental mode of a pendulum suspended by a filament is given
by\cite{Saulson_PRD}
\begin{equation}
\label{single filament dissipation dilution}
\Phi=\frac{1}{2}\sqrt{\frac{YI}{MgL^2}}\,\phi,
\end{equation}
where $\phi$ is the loss angle of the unloaded suspension filament,
$Y$ is the Young's modulus of the filament material, $I$ is the
cross-sectional moment of inertia, $M$ is the supported mass, $g$ is the
acceleration due to gravity, and $L$ is the filament length.  In
general, $\phi$ and hence $\Phi$ will be a function of frequency.
The coefficient of $\phi$ is called the dissipation dilution factor
and is the ratio of elastic energy (subject to dissipation) to the total
energy stored in the pendulum mode, which is predominantly gravitational
energy (non-dissipative).
The right hand side of this equation should be multiplied by two if the mass
is constrained from rotating in the plane of oscillation,
since bending then occurs both in the region where the filament leaves
the support and in the region where the filament leaves the mass.
If the test mass is suspended by $N$ filaments Eq.~\ref{single filament
dissipation dilution}
should be multiplied by $\sqrt{N}$.
For fibers of diameter $d_f$, and ribbons of thickness
$d_r$ and width $w$, we have
\begin{equation}
I=\left\{
\begin{array}{ll}
(\pi/64)\, d_f^4    &\qquad\mbox{{fibers,}}\\
(1/12)\,wd_r^3      &\qquad\mbox{{ribbons,}}
\end{array}\right.
\end{equation}
where subscripts $f$ refer to fibers  and subscripts $r$ to ribbons.
Rewriting Eq.~\ref{single filament dissipation dilution} using these
expressions for the cross-sectional moment of inertia, allowing multiple
filaments and assuming that the suspension constrains the filaments to bend
at both ends, we have
\begin{equation}
\label{dissipation dilution}
\Phi = \left\{
\begin{array}{ll}
\sqrt{\frac{YN_f \pi (d_f/2)^4}{4MgL^2}}\,\phi_f
    &\qquad\mbox{fibers,}\\
\sqrt{\frac{YN_r wd_r^3}{12MgL^2}}\,\phi_r
    &\qquad\mbox{{ribbons,}}
\end{array}\right.
\end{equation}
where $M$ is the suspended mass, $N_f$ is the number of suspension
fibers, $\phi_f$ is the loss angle of the unloaded fibers,
$N_r$ is the number of suspension ribbons, and $\phi_r$ is the
loss angle of the unloaded ribbons.
The limit to how much dissipation dilution we can obtain, and hence the
lower limit to the pendulum mode thermal noise, is set by the
values obtainable for the parameters in these equations.  They are
limited by a number of material and technological concerns, especially
by the value achievable for the loss angle $\phi_f$ or $\phi_r$.
This loss angle may depend on a number of factors including the
bulk material loss angle, surface loss, and the filament geometry.

If $\phi_f$ and $\phi_r$ were independent of filament thickness and roughly equal,
Eq.~\ref{dissipation dilution} would indicate that by using very thin
but wide ribbons one could obtain lower dissipation $\Phi$, and
hence less pendulum mode thermal noise, than by using fibers of similar
load bearing capacity. However, since surface loss becomes increasingly
important as the filament thickness is reduced, the enhanced dissipation dilution
obtainable using thin ribbons is moderated by an increase in $\phi_r$.


\section{Thermal Noise in the Presence of Surface Loss}

The loss angle for a sample, including surface loss, may be
expressed as\protect\cite{Andri}
\begin{equation}
\label{intrinsic dissipation}
\phi = \phi_{\mathrm{bulk}}(1+\mu\frac{d_s}{V/S}),
\end{equation}
where $\phi_{\mathrm{bulk}}$ is the loss angle of the bulk material,
$\mu$ is a geometrical factor and $d_s$ is the dissipation depth which
parametrizes the filament size at which surface loss becomes important.
The geometrical factor $\mu$ describes the emphasis placed on the condition
of the surface due to the sample geometry and mode of oscillation while the
dissipation depth $d_s$ describes the amount of surface damage and the depth to
which it penetrates.  Equation~\ref{intrinsic dissipation} serves to
define $d_s$, whose value for a given sample may be determined by experiment.
The geometrical factor is given by
\begin{equation}
\mu  = \frac{V}{S}\, \frac{\int_{\mathcal S} \epsilon^2({\vec r}) d^2r}
{\int_{\mathcal V} \epsilon^2(\vec{r}) d^3r},
\end{equation}
where ${\vec r}$ denotes a point in the sample, $\epsilon({\vec r})$ the
strain amplitude, $V$ is the volume of the sample, $S$ the surface area of
the sample,
${\mathcal V}$ is the set of points comprising the volume, and
${\mathcal S}$ is the set of points comprising the outer surface. For
transverse oscillations of fibers and ribbons we have
\begin{equation}
\label{mu's}
\mu=\left\{
\begin{array}{ll}
2       &\quad\mbox{fibers,} \\
(3+a)/(1+a)     &\quad\mbox{ribbons}
\end{array}\right.
\end{equation}
where $a$ is the aspect ratio of the combined ribbons, $a\equiv d_r/W$
with $W\equiv N_r w$ being the total combined width of the ribbons.

Experiments suggest that $\phi_{\mathrm{bulk}}$ is approximately
constant over the frequency range of interest for
LIGO\cite{Lunin,Andri,Bill}. For simplicity, we will assume
$\phi_{\mathrm{bulk}}$ to be constant. Substituting
Eqs.~\ref{intrinsic dissipation}~and~\ref{mu's} into
Eq.~\ref{dissipation dilution} we obtain
\begin{equation}
\label{dd with surface effect}
\Phi= \left\{
\begin{array}{ll}
\sqrt{{Y}/{16\sigma L^2}}\,(d_f+8d_s)\phi_{\mathrm{bulk}}
    &\quad\mbox{{fibers,}}
\vspace{3pt}\\
\sqrt{{Y}/{12\sigma L^2}}\,(d_r+(6+2a)d_s)\phi_{\mathrm{bulk}}
    &\quad\mbox{{ribbons,}}
\end{array}\right.
\end{equation}
where $\sigma$ is the filament stress.
In both cases, the first term is the traditional expression for
dissipation dilution, while the term involving $d_s$ represents a reduction
of the dilution due to
the increasing importance of surface loss as the filament thickness is
decreased. For
very thin filaments the term involving $d_s$ dominates and the loss angle
becomes independent of the filament thickness.

From the fluctuation-dissipation theorem~\cite{Callen}, we find
the power spectrum of the pendulum mode thermal fluctuations above
the pendulum mode resonance, at angular frequency $\omega$:
\begin{equation}
\label{FDT pend}
x^2(\omega)  = \frac{4k_BTg}{ML'\omega^5}\,{\Phi(\omega)},
\end{equation}
where $\omega \stackrel{>}{_\sim} \sqrt{g/L'}$, $T$ is the temperature of the suspending filaments and $L'$
is the radius of the arc traced out by the center of mass during
pendulum mode oscillation.  For convenience, we will take $L'\approx L$.
Inserting $\Phi$ from Eq.~\ref{dd with surface effect}, and including
the contribution from thermoelastic damping
we have the expression for the pendulum mode thermal noise:
\begin{equation}
\label{pend fluct}
\begin{array}{l}
x^2(\omega)  =   \vspace{5pt}
 \left\{\hspace{-4pt}
\begin{array}{ll}
\frac{4k_BTg}{ML^2\omega^5}\,\sqrt{\frac{Y}{16\sigma}}\,
\Big[d_f\left(\phi_{\mathrm{bulk}}+\phi_{\mathrm{th}}\right)+8d_s\phi_{\mathrm{bulk}}\Big]
&\quad\mbox{{fibers,}} \vspace{3pt}\\
\frac{4k_BTg}{ML^2\omega^5}\,\sqrt{\frac{Y}{12\sigma}}\,\Big[d_r\left(
\phi_{\mathrm{bulk}} +\phi_{\mathrm{th}}\right)+(6+2a)d_s\phi_{\mathrm{bulk}}\Big]
&\quad\mbox{{ribbons.}}
\end{array}  \right.
\end{array}
\end{equation}
The thermoelastic damping term $\phi_{\mathrm{th}}$ is given by\cite{Zener}
\begin{equation}
\label{thermoelastic damping}
\phi_{\mathrm{th}}=\frac{Y\alpha^2T}{C}\frac{\omega\tau_d}{1+\omega^2\tau_d^2},
\vspace{3pt}
\quad \tau_d=\left\{\hspace{-4pt}
\begin{array}{ll}
d_f^2/13.55D    & \quad\mbox{fibers,} \vspace{3pt}\\
d_r^2/\pi^2D    & \quad\mbox{ribbons,}
\end{array} \right.
\end{equation}
where $\alpha$ is the thermal expansion coefficient of the filament
material, $C$ is the heat capacity per unit volume, and $D$ is the
thermal diffusion coefficient.


\section{Advanced Interferometers}
\label{Advanced Interferometers} Using Eq.~\ref{pend fluct}, we
can now make estimates for the level of pendulum mode thermal
noise achievable in advanced interferometers and investigate the
dependence on filament thickness and geometry. Using the results
of previous experiments and design studies, most of the
parameters can be well bounded with some reasonable assumptions.
From these parameters, we can obtain upper and lower bounds on
the pendulum mode thermal noise at a given frequency as a
function of ribbon thickness.  This analysis assumes that losses
extrinsic to the filaments (e.g. recoil of the suspending
structure or lossy filament-to-test-mass bonds) have been made
negligible.

The achievable fiber diameter depends on the achievable stress
$\sigma$ and on the mass $M$.  The fiber diameter in
Eq.~\ref{pend fluct} can be replaced by
\begin{equation}
\label{fiber_diameter}
d_f=\sqrt{4Mg/\pi N_f\sigma} .
\end{equation}
The remaining parameters $M$,$L$,$N_f$, $\sigma$, and $d_s$ are
independent and the achievable pendulum mode thermal noise
depends on the bounds established for these parameters.

It is clear from  Eq.~\ref{pend fluct} that an efficient way of
minimizing the thermal noise is to make the length of the
suspension as large as possible. However, the value of $L$ is
bounded above by the minimum allowable spacing $f_{\mathrm{min}}$
of the violin mode frequencies. The frequency spacing must be
kept above about 300~Hz to allow reasonably large intervals of
the spectrum to be free of violin modes. The frequency spacing
limited $L$ is
\begin{equation}
\label{L_max}
L=\frac{1}{2f_{\mathrm{min}}}\sqrt{\frac{\sigma}{\rho}}
\end{equation}
where $f_{\mathrm{min}}$ is the minimum allowable spacing of the
violin modes. The range of possible lengths is determined by the
range of possible stress to which the filaments will be subject.
This in turn depends on the breaking strengths achievable for
fused silica filaments. Many measurements have been reported on
the breaking strength of fibers manufactured from naturally
occurring, and synthetic, vitreous
silica\cite{Russian_book,Proctor}. Little is known about the
strength of ribbons, though one is tempted to assume their
strengths are similar. Values reported for the breaking strength
of fibers in tension at room temperature vary greatly depending
on the condition of the fibers~\cite{Ernsberger}, but strengths on
the order of  several gigapascals at room temperature, in fibers
with diameters as large as 1~mm have been
reported~\cite{Hillig,Morley,Perugia}. By assuming that the
filaments are only loaded to a fraction of their breaking
strength we assign the range of possible stress to which the
filaments will be subject as
$0.1~\mathrm{GPa}\nobreak\leq\nobreak\sigma\nobreak\leq\nobreak
1.0~\mathrm{GPa}$. Substituting these values into Eq.~\ref{L_max}
we obtain the range of possible lengths
$0.36~\mathrm{m}\nobreak\leq\nobreak
L\nobreak\leq\nobreak1.1~\mathrm{m}$. In principle, the physical
design of the suspension also places an upper limit on the
length, but ultimately this limit is likely to be less stringent
than that due to the frequency spacing.

For the number of filaments we will choose $N_f=N_r=4$. This
reflects the most likely choice for the suspensions in advanced
detectors which require ``marionette'' control of the test
masses~\cite{white_paper}. Analytically, the number of ribbons
does not enter into the calculation as, for a given stress
$\sigma$ and ribbon thickness $d_r$, only the total  combined
width of the ribbons $W$ is fixed.

In order to avoid excessive radiation pressure noise in LIGO II,
the suspended test masses must have a mass $M$ of 30~kg. However,
if this is not feasible, the LIGO I mass of 10~kg can be used as
a fall-back.  We take 10~kg~$<M<$~100~kg to allow for possible
advanced designs that utilize even larger
masses~\cite{white_paper}.

For the bulk material dissipation in fused silica, Gretarsson and
Harry have measured
$\phi_{\mathrm{bulk}}=3.3\pm0.3\times10^{-8}$\cite{Andri}. Others
have measured similar  values for the bulk material dissipation in
samples of different geometry~\cite{Lunin,Bill,Litten,Geppo:talk}.
We shall adopt the relatively reliable upper limit
$\phi_{\mathrm{bulk}}=3.6\times 10^{-8}$ and somewhat uncertainly
set the lower limit as $\phi_{\mathrm{bulk}}=2.5\times 10^{-8}$

Finally, for untreated fused silica fibers drawn in a natural gas
flame, $d_s=180\pm20~\mu$m has been  measured~\cite{Andri}. It
should be noted that the factors resulting in a given quality of
fiber surface are not well quantified.  Fibers pulled from silica
rods with different initial surface conditions, or fibers drawn
using a different production method, may have a different
dissipation depth or surface loss than that found in the
measurements above. The geometry of a filament could also have
some effect on the quality of the surface layer, e.g. through
different cooling stresses during fabrication, and our assumption
that fused silica ribbons have the same surface properties as
fused silica fibers has not been tested.  However, given these
assumptions we set an upper bound of $d_s=200~\mu$m, which should
be reliable. To estimate a lower bound for $d_s$ we will use a
$Q$-measurement of a ribbon of thickness $50~\mu$m, made of
natural fused quartz~\cite{Rowan}. One mode of this ribbon showed
a $Q$ much higher than others. After subtracting the loss due to
thermoelastic damping (Eq.~\ref{thermoelastic damping}) and
assuming the remaining loss is mainly surface loss, the
equivalent $d_s$ for a similarly limited fused silica ribbon can
be estimated at $30~\mu$m.  We, therefore, set the range of
possible dissipation depths at 30~$\mu$m$\nobreak\,<d_s<200~\mu$m.

\begin{figure}
\begin{center}
\epsfxsize=15cm
\leavevmode
\epsfbox{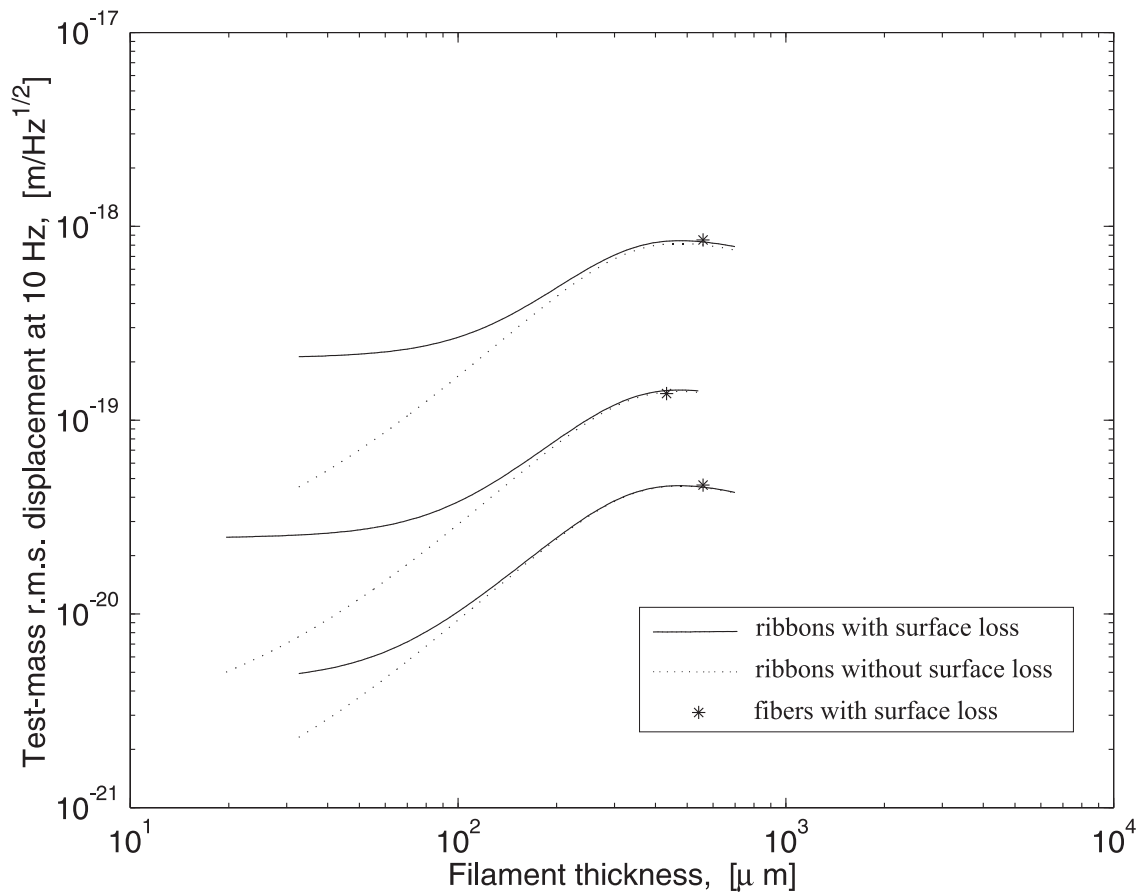}
\end{center}
\caption{Solid lines: Worst case (column~1 in
Table~\ref{estimates}), most probable (column~2 in
Table~\ref{estimates}) and best case (column~3 in
Table~\ref{estimates}) estimates for the level of pendulum mode
thermal noise for ribbons vs. ribbon thickness. Dashed lines:
Same, but neglecting surface loss. Asterisks: Worst case
(column~1 in Table~\ref{estimates}), most probable (column~2 in
Table~\ref{estimates}) and best case (column~3 in
Table~\ref{estimates}) estimates for the level of pendulum mode
thermal noise in 4-fiber suspensions with maximally loaded
fibers. In the case of ribbons, the minimum thickness is set by
requiring the total combined width of the ribbons $W$ be less
than or equal to 3 cm. The maximum ribbon thickness is set by
requiring $d_r/W \le 0.5$.}

\label{xvsd}
\end{figure}

Table~\ref{estimates} summarizes the best and worst case
estimates for the parameters, and also a ``best guess'' for the
most probable values.  Figure~\ref{xvsd} shows the levels of
pendulum mode thermal noise at 10~Hz as a function of filament
thickness for each of the three sets of estimates of the
parameters. The graphs for ribbon filaments all show a maximum
around $400~\mu$m. This is the thermoelastic damping peak, which
for all but the most optimistic case must be avoided if the
desired levels of pendulum mode thermal noise are to be achieved.
Figure~\ref{xvsd} also shows the value of using ribbons rather
than fibers as suspension filaments. Since the diameter of fibers
cannot be independently reduced, the pendulum mode thermal noise
is dominated by thermoelastic damping.  The use of ribbons allows
us to evade this problem by residing in the surface loss limited
regime.  To evade the thermoelastic regime and to obtain better
dissipation dilution one might be tempted to use the thinnest
ribbons possible.  However, at small thicknesses, below the
thermoelastic peak, the graphs begin to level off.  This is due
to surface loss which sets a minimum to achievable pendulum mode
thermal noise of
\begin{equation}
x^2(\omega)_{\mathrm{min}} =
\frac{24k_BTg}{ML^2\omega^5}\,\sqrt{\frac{Y}{12\sigma}}\,d_s\phi_{\mathrm{bulk}},
\label{x_min}
\end{equation}
It is clear from the plots that, for all but the most optimistic
values of $d_s$, reducing the ribbon thickness below about
$50~\mu$m (corresponding to individual ribbon widths $w$ of 5~mm,
3~mm, and 5~mm for the three cases) does not result in
significant reductions of pendulum mode thermal noise. To satisfy
LIGO II requirements, the pendulum mode thermal noise at 10~Hz
should be less than about $2\times
10^{-19}~\mathrm{m/\sqrt{Hz}}$~\cite{white_paper}.  Even in the
presence of surface loss, only the worst case scenario does not
achieve this level. In the most probable case, pendulum mode
thermal noise will be lower than other noise sources at 10~Hz
(radiation pressure noise, fused silica internal mode thermal
noise), provided the ribbon thickness is kept below the
thermoelastic regime. In the most optimistic case, other noise
sources dominate the total noise at 10~Hz regardless of ribbon
thickness. The most probable estimate for fiber suspensions gives
pendulum mode thermal noise that is just acceptable for LIGO II.
If there are unforeseen problems with ribbon suspensions, fiber
suspensions may still prove an acceptable alternative. We
reiterate that in the comparison of fibers and ribbons, we have
assumed that the breaking strength of fibers is not significantly
greater than that of ribbons of equal cross-sectional area; we
have also assigned them identical surface properties.  Further
research is required to test these assumptions.  Research is
continuing on ribbon suspensions within the LIGO research
community, and additional emphasis on surface properties and
breaking strength is warranted.


\section{Comparison of low frequency noise sources in LIGO II}

While studying the pendulum mode thermal noise at 10~Hz is a good way to
gain insight into the effect of the different physical parameters on the
level of this source of noise, the comparison with other sources of noise
must be done over the entire range of relevant frequencies.

For clarity, we now specialize to a single set of values for the
physical parameters of the suspension filaments--those proposed
for LIGO II~\cite{white_paper}.  These parameters are shown in
the last column of Table~\ref{estimates}. With a four ribbon
suspension, it follows from the proposed stress that each ribbon
should have a cross-sectional area of $5.5 \times
10^{-7}~\mathrm{m}^{2}$, giving a width of 5.5~mm for the
$100~\mu$m ribbon thickness proposed.   In general, the values for
all these parameters fall between the worst case and most probable
case scenarios.  As such, the LIGO II proposal is fairly
conservative, and better noise performance may be achieved.

The LIGO II proposal does not, however, specify a value for $d_s$.
In keeping with the conservative spirit of the other parameters,
we choose $d_s = 180~\mu$m.   This falls between the worst case
and most probable case and corresponds to the measured value in
fibers without strict handling requirements~\cite{Andri}.

\begin{figure}
\begin{center}
\epsfxsize=15cm \leavevmode \epsfbox{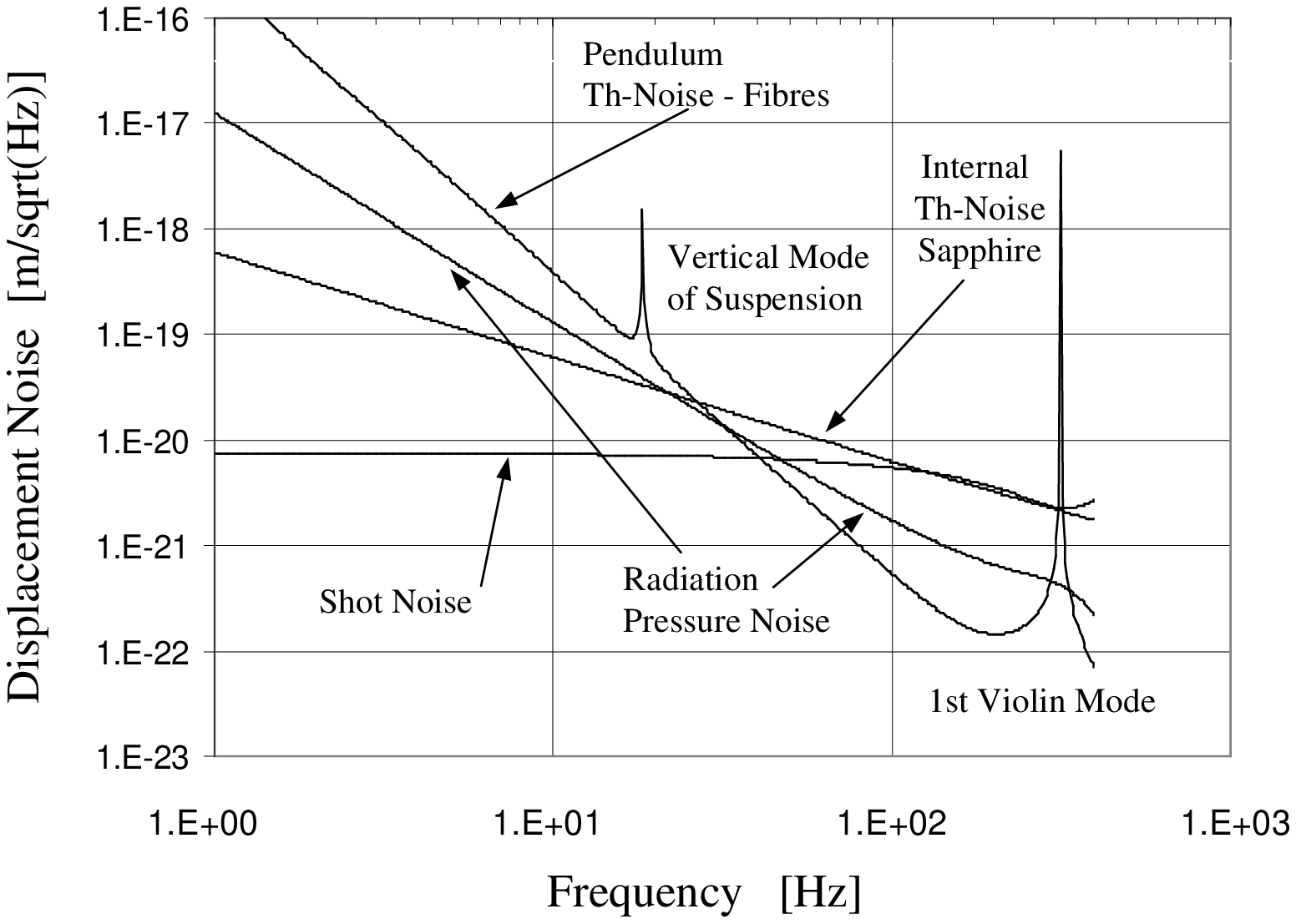}
\end{center}
\caption{Frequency domain comparison of pendulum mode thermal
noise of a fiber suspension with other low frequency noise sources
in LIGO II. The assumptions in the LIGO II white paper concerning
the output optics have been slightly modified to reflect the
addition of thermoelastic damping in the test masses. The noise is
expressed as motion of the face of a single test mass.}
\label{noise_fibers}
\end{figure}

\begin{figure}
\begin{center}
\epsfxsize=15cm \leavevmode \epsfbox{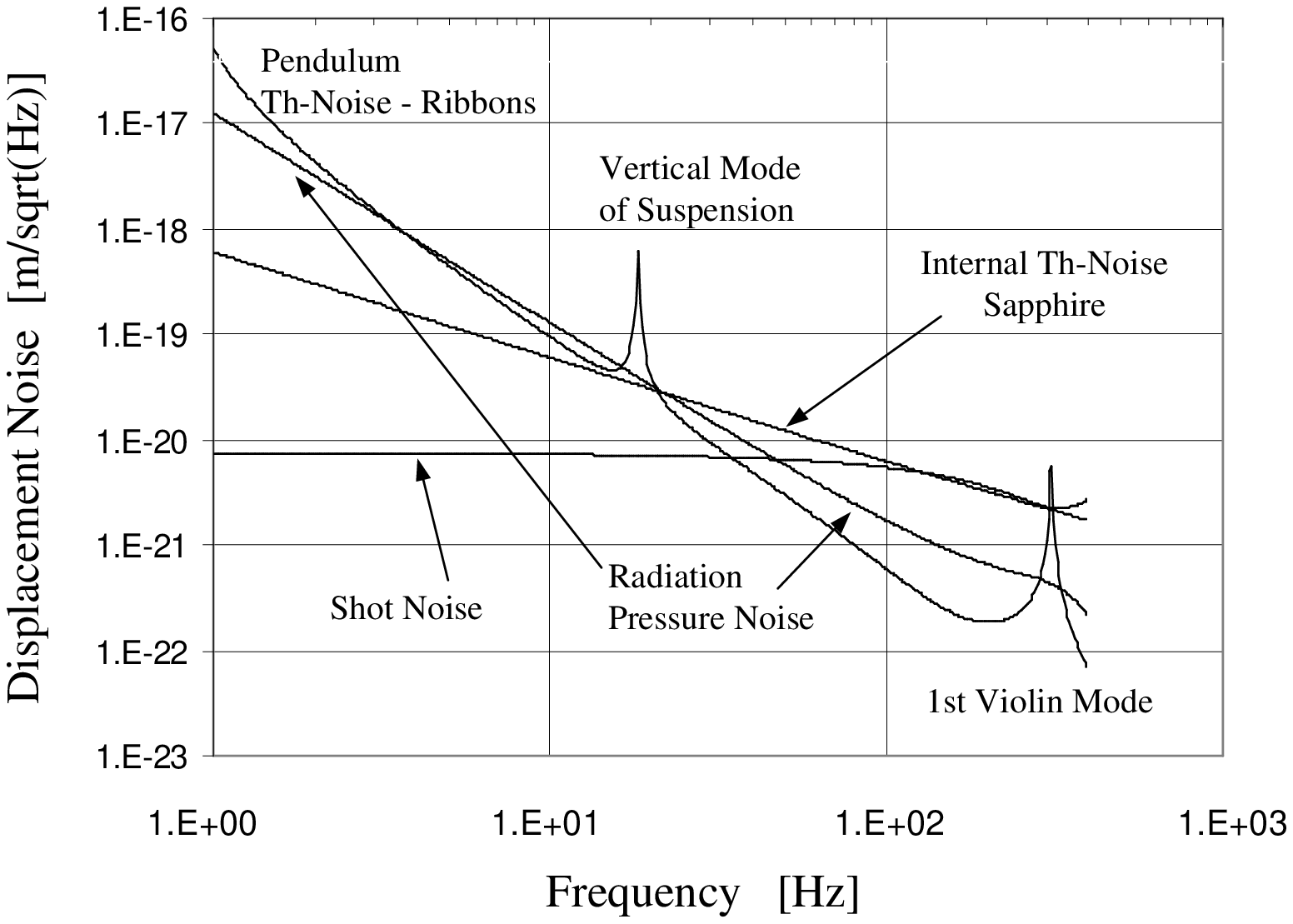}
\end{center}
\caption{Frequency domain comparison of pendulum mode thermal
noise of a ribbon suspension with other low frequency noise
sources in LIGO II. The assumptions in the LIGO II white paper
concerning the output optics have been slightly modified to
reflect the addition of thermoelastic damping in the test masses.
The noise is expressed as motion of the face of a single test
mass.} \label{noise_ribbons}
\end{figure}

Figure~\ref{noise_fibers} shows the pendulum mode thermal noise of
a fiber suspension in relation to estimates for the other sources
of noise in the interferometer.  Figure~\ref{noise_ribbons} shows
the same comparison for the pendulum mode thermal noise of a
ribbon suspension~\cite{Braginsky_note}.  Note that the noise due
to radiation pressure in the LIGO II design is greater than the
pendulum mode thermal noise for the ribbon suspension. For
greater low frequency sensitivity, radiation pressure can always
be reduced by lowering the amount of laser power at the beam
splitter.  This could reduce the noise to the pendulum mode
thermal noise level in the low frequency band but will increase
the amount of laser shot noise at higher frequencies. From
figures~\ref{noise_fibers}~and~\ref{noise_ribbons} it is clear
that while a ribbon suspension leads to lower pendulum mode
thermal noise than a fiber suspension, the pendulum mode thermal
noise for the fiber suspension is still comparable to radiation
pressure noise in the relevant frequency band. If there are
unforeseen problems with ribbons (buckling, lower strength,
etc.), fibers do provide an acceptable, if less attractive,
alternative.


\section{Acknowledgments}

We would like to thank our colleagues at the University of
Glasgow, at Stanford University, and throughout the gravitational
wave community for their interest in this work.  Additional
thanks to Ken Strain for his help with LIGO II parameters beyond
the white paper as well as Gabriela Gonzalez, Gary Sanders, David
Tanner, and Rai Weiss for their comments. This work was supported
by Syracuse University, U.S. National Science Foundation Grants
No. PHY-9602157 and No. PHY-9630172, the University of Glasgow,
and PPARC.




\pagebreak[4]
\begin{table}
\begin{tabular}{lllll}
 &worst case     &most probable      &best case      &LIGO II
proposal \\ \hline
$\phi_{\mathrm{bulk}}$      &$3.6\times 10^{-8}$    &$3.3\times 10^{-8}$    &$2.5\times 10^{-8}$    &$3.3 \times 10^{-8}$\\
$d_s$               &$200~\mu$m     &$100~\mu$m     &$30~\mu$m      &$180~\mu$m$^{*}$ \\
$\sigma$            &0.1 GPa        &0.5 GPa        &1.0 GPa        &0.13~GPa \\
$M$             &10 kg          &30 kg          &100 kg         &30~kg \\
$L$             &0.35 m         &0.79 m         &1.1 m          &0.39~m

\end{tabular}
\caption{Estimates for the parameters determining the pendulum
mode thermal noise. \newline $^{*}$not specified in proposal}
\label{estimates}
\end{table}

\pagebreak[4]

\pagebreak[4]

\pagebreak[4]

\end{document}